# Spectacular doping dependence of interlayer exchange and other results on spin waves in bilayer manganites

T.G.Perring[1], D.T.Adroja[1], G.Chaboussant[1], G.Aeppli[2], T. Kimura[3], Y.Tokura[3,4]

[1]*ISIS Facility, CLRC Rutherford Appleton Laboratory, Didcot, Oxon OX11 0QX, United Kingdom*
[2]*NEC Research Institute, 4 Independence Way, Princeton, New Jersey 08540*
[3]*Department of Applied Physics, University of Tokyo, Tokyo 113-8656, Japan*
[4]*Joint Research Center for Atom Technology (JRCAT), Tsukuba 305-0046, Japan*

(Version date: 9 May 2001)

We report the measurement of spin waves in the bilayer colossal magnetoresistive manganites $La_{2-2x}Sr_{1+2x}Mn_2O_7$ with $x$=0.30, 0.35 and 0.40. For $x$=0.35 and 0.40 the entire acoustic and optic dispersion relations are reasonably well described by those for a bilayer Heisenberg Hamiltonian with nearest neighbour exchange only. The in-plane coupling is only weakly dependent on $x$, but the coupling between the planes of a bilayer changes by a factor of four. The results directly reveal the change from mixed $d_{3z^2-r^2}$ and $d_{x^2-y^2}$ character of the orbitals to mostly $d_{x^2-y^2}$ character with increasing hole concentration.

The cubic manganese perovskites $RE_{1-x}A_xMnO_3$ (RE=La, Nd, Pr, A=Sr, Ca, Pb, for example) have attracted attention because of their rich and fascinating physical properties. Depending on the composition these can include not only changes in resistivity of several orders of magnitude in an applied field of a few Tesla, but also magnetic field, electric field-, photon- and strain-induced insulator-to-metal phase transitions [1]. The starting point to understand the manganites is double exchange (DE) theory [2,3] in which electrons in partially filled bands with $e_g$ character hop between adjacent Mn ions, each of which has a spin $S$=3/2 magnetic moment arising from local $t_{2g}$ orbital occupancy. Strong intrasite exchange, $J_H$~2 eV favours hopping between parallel Mn spins and therefore ferromagnetism and metallic conductivity. DE alone cannot explain the physics of the manganites, however, and there is a wealth of experimental data pointing towards lattice distortions which localise carriers to form polarons in the paramagnetic phase [4].

Naturally layered manganites [5] allow the influence of another control parameter, that of dimensionality to be explored. In addition, they provide the venue for an object not found elsewhere in nature, namely a two-dimensional, fully spin-polarised electron liquid. $La_{2-2x}Sr_{1+2x}Mn_2O_7$ consists of bilayer slices of $MnO_6$ octahedra taken from the cubic compound, separated by insulating $(La,Sr)_2O_2$ layers that serve to largely decouple the bilayers both electronically and magnetically (Fig. 1(a)). The influence of reduced dimensionality is seen in the enhanced CMR near the three-dimensional ordering temperature $T_C$ as compared to the cubic (La,Sr)MnO$_3$ analogue, albeit at the cost of reduced $T_C$~100–120 K [5].

Surprisingly, there are very few results for the spin dynamics in bilayer manganites [6-8]. Published work is confined to just $x$=0.4, with only one preliminary report of measurements to the zone boundary [8], and no quantitative analysis of the spin wave lifetimes as a function of momentum. Here we present the results of a study as a function of hole doping in the range $x$=0.30–0.40. We show that the inter-plane (but still intra-bilayer) coupling shows a remarkable sensitivity to hole doping, which can be explained by changing $d_{3z^2-r^2}$ orbital occupancy, and that for $x$=0.35 and $x$=0.40 the spin waves throughout the Brillouin zone can be understood solely within the framework of the DE model.

Our neutron scattering data were taken for single crystals of La$_{2-2x}$Sr$_{1+2x}$Mn$_2$O$_7$ with masses 0.7, 2.9 and 1.9g for $x$=0.30, 0.35 and 0.40 respectively. Neutron diffraction confirmed that the moments in the ordered phases are ferromagnetically ordered within the bilayers [9-12]. For $x$=0.35 and 0.40 the bilayers have weak ferromagnetic coupling to neighbouring bilayers, while for $x$=0.30 the coupling is antiferromagnetic. In this paper, we are concerned only with the *intra*-bilayer coupling; in fact the *inter*- bilayer coupling is unmeasurable in these experiments, being 100 times smaller [13]. The spin wave measurements were performed at $T \approx 10$ K $\ll T_C$ or $T_N$ on the HET and MARI spectrometers of the ISIS pulsed neutron source at the Rutherford Appleton Laboratory, UK. The spectrometers are well suited to study spin waves in the manganites, which typically have a bandwidth of 100 meV, because of their low background and high flux of epithermal neutrons [14]. From just a few choices of incident energy and crystal orientation the entire dispersion relation along the major symmetry directions could be mapped out for $x$=0.35 and $x$=0.4. With the smaller $x$=0.3 crystal data could be obtained only near the zone centre.

To analyse the data we considered the Heisenberg Hamiltonian with coupling $J_{ij}$ between localised spins at sites $i$ and $j$, $H = -\sum_{<i,j>} J_{ij} \mathbf{S}_i \cdot \mathbf{S}_j$. The simplest form of the Hamiltonian is when there are only nearest neighbour interactions: $J_\parallel$ along Mn-O-Mn bonds in the plane and $J_\perp$ along Mn-O-Mn bonds that connect the two planes of a bilayer. In this instance, the acoustic and optic spin wave dispersion relations are $\hbar\omega_{ac}(Q) = 4SJ_\parallel (\sin^2(Q_x a/2) + \sin^2(Q_y a/2))$ and $\hbar\omega_{op}(Q) = \hbar\omega_{ac}(Q) + 2SJ_\perp$, and the neutron scattering cross section per Mn site is

$$\frac{d^2\sigma}{d\Omega d\omega} = \left|\frac{k_f}{k_i}\right| (\gamma_N r_0)^2 \frac{S_{eff}}{2} |F(Q)|^2 \left(1 + (Q_z/Q)^2\right) \times \left(\cos^2(Q_z \Delta z/2) S^{ac}(Q,\omega) + \sin^2(Q_z \Delta z/2) S^{op}(Q,\omega)\right)$$

where $S^{ac,op}(Q,\omega) = (n_\omega + 1)\delta(\omega - \omega_{ac,op}(Q)) + n_\omega \delta(\omega + \omega_{ac,op}(Q))$, with $n_\omega = \left(1 - e^{-\beta\hbar|\omega|}\right)^{-1}$. Here $F(Q)$ is the magnetic form factor and $(\gamma_N r_0)^2$ is 291 mbarns/steradian. The possibility of finite spin-wave lifetimes was allowed for by replacing the delta-function response with that for a damped simple harmonic oscillator with inverse lifetime $\gamma$. To place points on the acoustic spin wave dispersion relation, each cut with integer $Q_z \Delta z/2\pi$ was independently fitted to the convolution of the neutron scattering cross-section and the instrument resolution. The optic spin wave dispersion relation was established by using the previously determined acoustic dispersion to independently derive an effective $J_\perp$ for each spin wave in cuts with half-integer $Q_z\Delta z/2\pi$. In this fashion, each acoustic and optic spin wave peak was independently assigned a position on the dispersion relation, an intensity and a lifetime. The results are summarised in Figs. 2 and 3.

We consider first the dispersion relations for $x$=0.35 and $x$=0.40, shown in Fig.2. The principal features are that well-defined spin waves exist to at least 70meV, and that there is no softening of the branches near the zone boundary. The solid lines show the best fit to the data for the nearest neighbour Heisenberg Hamiltonian. Except near the (½,0) zone boundary for the optic branch, the model provides a reasonably good description of the entire spin wave dispersion, with $SJ_\parallel$=9.6±0.1meV and $SJ_\perp$=2.9±0.1meV for $x$=0.40, and with $SJ_\parallel$=9.3±0.1meV and $SJ_\perp$=5.7±0.2meV for $x$=0.35. The relative dispersion of the optic mode along ($h$,0) with respect to the acoustic mode can be accounted for by a small second neighbour interaction $SJ_{a,0,a} \approx$ -0.8meV that connects the two planes of a bilayer along the direction [a,0,a] in the real space lattice. (In the same notation $J_\parallel$ is $J_{a,0,0}$ and $J_\perp$ is $J_{0,0,a}$). The corresponding dispersion relations are shown by the dotted lines in Fig.2. The other second nearest neighbour interactions, $J_{a,a,0}$ and $J_{2a,0,0}$, do not improve the fit, and $J_{a,a,a}$, which yields a similar improvement as $J_{a,0,a}$, we dismissed on the basis that it requires a carrier to make *three* hops along Mn-O-Mn bonds and is therefore expected to be weaker still. We note that the values of $SJ_\parallel$ and $SJ_\perp$ for $x$=0.4 are in good agreement with reports from other groups [6,7], but that those data were confined to $\hbar\omega \leq$25meV, and were therefore unable to test the applicability of the nnHFM model.

The spin wave dispersion for the DE model is precisely that for the nnHFM model in the $J_H = \infty$ RPA limit [15,16]. Treatment to second order in a 1/S expansion for the 2D DE model (but still $J_H = \infty$ for a single band model) [17] yields a reduction of about 50% in the stiffness $D$ (defined by $\hbar\omega = Dq^2$ in the small $q$ limit), but in fact $\hbar\omega$ at $q$=(½,0) is 10-20% greater than for the nnHFM with that reduced stiffness. ( ref.17 Fig.1). A similar reduction of 35% in overall energy scale, with a small hardening towards zone boundary, was found in 3D for $x$=0.30 using a variational approach that further allowed for finite $J_H$ =W/2 [18]. Data from band structure calculations (majority carrier full band width $W \approx$2-3eV, $J_H \approx$1-1.5 eV $\approx W/2$) [19] in conjunction with results from the *1/S* expansion yield $\hbar\omega$(½,0)≈20-40 meV, in good quantitative agreement with our data. Remarkably, a simple one-band DE model provides an explanation for both the energy scale and near-nnHFM dispersion relation for the data.

For comparison with our less comprehensive data for $x$=0.30, we calculated the spin wave stiffness and optic gap from the fitted exchange constants, and used the $q\rightarrow$0 limits of the dispersion relations for the nnHFM to obtain effective nearest neighbour exchange constants $SJ_\parallel^{\text{eff}}$ and $SJ_\perp^{\text{eff}}$. These are summarized in Fig. 4(a). $SJ_\parallel^{\text{eff}}$ is only weakly dependent on the hole doping $x$, and is very similar to values obtained in the 3D manganites [20-24]. In contrast, $SJ_\perp^{\text{eff}}$ is remarkably sensitive to the hole doping, with $J_\perp^{\text{eff}}/J_\parallel^{\text{eff}}$ changing from 1.1 to 0.25 over the range $x$=0.30-0.40. The results reveal the changing character of the $e_g$ orbital occupancy of the Mn ions with doping. Ferromagnetic exchange between the planes arises from orbital overlap of O $p_z$ and Mn $d_{3z^2-r^2}$ orbitals, whereas ferromagnetic exchange in a plane is due to overlap of O $p_x$ or O $p_y$ and Mn $d_{x^2-y^2}$ orbitals (Fig.1). Increasing $SJ_\perp^{\text{eff}}$ therefore directly implies increasing occupancy of $d_{3z^2-r^2}$ orbitals in preference to $d_{x^2-y^2}$ orbitals as the hole doping decreases, or equivalently as the $e_g$ electron number density increases. The conclusion can be understood using the results of diffraction studies [10-12], which show that the MnO$_6$ octahedra expand significantly along the *c*-axis as the hole doping is reduced, as reflected in the increasing ratio of the mean apical Mn-O to equatorial Mn-O distances, $\Delta_{JT}$ (Fig.4(c)). Elongation along the *c*-axis lowers the energy of the $d_{3z^2-r^2}$ orbital with respect to the $d_{x^2-y^2}$ orbital, in accord with our conclusion. Indirect evidence of the preferential filling of the $d_{3z^2-r^2}$ band as hole doping is reduced has come from anisotropic magnetostriction [25] and optical conductivity data [26]. Our measurements of the interplane exchange directly demonstrate the hypothesis.

We now turn to the lifetime of the spin waves. Fig.3(a) shows γ as a function of excitation energy along (*h*,0) for *x*=0.40. For both the acoustic and optic modes, γ increases approximately linearly with excitation energy out to the zone boundary, with the damping a substantial fraction of the excitation energy $-\gamma(q)/\hbar\omega(q)$ =0.33±0.02 and 0.46±0.02 for the acoustic and optic modes respectively. In particular, there is no steep increase where the spin wave branch crosses an optic phonon mode at ≈20 meV [26]. (Note: there is a small deviation of the lowest energy optic modes from the linear relation. At the zone centre γ was found to be 3.8±0.7meV, the same as 4.0±0.2meV in [7].) We observed similar behaviour – linear $\gamma(q)/\hbar\omega(q)$ to the zone boundary for both acoustic and optic modes – for *x*=0.35 ($\gamma(q)/\hbar\omega(q)$ =0.24±0.01 and 0.38±0.01 respectively). The important aspects of the data are that the slopes are of order unity, and that the damping rates are much larger than $k_B T \approx 1$ meV throughout most of the Brillouin zone. Spin wave broadening can arise from static or dynamical fluctuations. The first would translate into quenched disorder which would cause the exchange constants to belong to distributions of width $\Delta J_\parallel$ and $\Delta J_\perp$, immediately leading to spin wave decay rates $\gamma(q)=\Delta\hbar\omega(q)$. Because the measured slopes for La$_{2-2x}$Sr$_{1+2x}$Mn$_2$O$_7$ are of order unity, this would imply that the site to site fluctuations in *J* would be of order *J* itself, something which might well yield carrier localization, in contradiction with the low temperature metallicity of our samples. This leaves us to consider dynamical damping mechanisms, although we do not exclude a small static contribution to the ratio $\gamma(q)/\hbar\omega(q)$. There are several possible channels: (i) magnon-magnon scattering, (ii) magnon-phonon coupling, (iii) coupled orbital-lattice fluctuations, and (iv) magnon-electron scattering. Case (i) has not been considered for DE, but experiment and theory for the planar nnHFM suggest that it is unlikely to explain the heavy damping. K$_2$CuF$_4$ provides an excellent realisation of the planar nnHFM for *S* = ½, and nowhere in the Brillouin zone was $\gamma(q)/\hbar\omega(q)$ found to exceed 0.02 for $T/J < 0.25$ [27]. For $q\rightarrow 0$ the leading order expression for γ is linear in excitation energy, $\gamma(q)/\hbar\omega(q) = (1/2\pi)(T/JS^2)^2$ [28], which when evaluated at $T = 10$ K with $SJ_\parallel^{eff}$ and $S = (4-x)/2$ yields ≈1.6×10$^{-3}$, ~100 times smaller than our measurements. Case (ii) may dampen the spin waves without softening the dispersion, via modulation of the DE exchange as the Mn-O-Mn distance oscillates [29]. For $\hbar\omega(q)$ greater than that of a lattice mode associated with distortion of the MnO$_6$ octahedra, $\Omega$, scattering of a spin wave into a phonon and lower energy spin wave is kinematically allowed. In 2D, γ is expected to show a step-function increase at $\Omega$. Significant broadening at optic phonon frequencies has been observed in La$_{0.7}$Ca$_{0.3}$MnO$_3$ [21]. However, no step in γ is observed in our data, indicating that another mechanism must be sought [30]. The effect of (iii), coupled orbital-lattice fluctuations, softening the spin-wave branch near the zone boundary has been treated quantitatively [31], but the effect on the damping of spin waves has not been considered. However, the absence of softening of the dispersion relation suggests that this mechanism is probably unimportant. Golosov considered Case (iv), magnon-electron scattering, at $T=0$ for DE [17]. Band width $W=2.5$ eV yields $\gamma(½,0)/\hbar\omega(½,0) = 1-4\times10^{-2}$ for the $x$ considered, ~10 times too small, but elsewhere on the zone boundary $\gamma/\hbar\omega$ can reach 0.15. This magnitude and the sensitivity of $\gamma(q)$ to precise details of the band structure suggests the plausibility of this mechanism to explain our results. Numerical simulations in 1D [32] support this conclusion. Furukawa [33] has pointed out the existence of a strongly $T$-dependent magnon-electron contribution to $\gamma$ which is also linear in $\omega$, as we have observed. However, this contribution vanishes in the $T\rightarrow0$ limit.

Finally, we consider the absolute value of the scattering cross-section. Figs.3(c-d) show that $S_{eff}$ is nearly constant along $(h,0)$ for $x=0.4$, and $x=0.35$ shows similar behaviour. $S_{eff}(q)$ is constant for the nnHFM, with deviations less than 5% from the nnHFM value for the DE model [18]. Our value of $S_{eff}=1.0$–$1.2$ for both $x=0.35$ and $0.40$ is somewhat smaller than $(4-x)/2\approx1.8$. However, systematic error arising from the calibration method (normalisation by vanadium standard) can easily be 30%. Interestingly, we have $S_{eff}=1.7\pm0.2$ for $x=0.30$.

In summary, we have measured the doping-dependent spin waves in a two-dimensional manganite. The three major findings are (i) the strong doping dependence of the inter-plane (but intra-bilayer) magnetic exchange, which directly demonstrates the preferential filling of the $d_{3z^2-r^2}$ band with decreasing hole concentration, (ii) the low temperature spin wave dispersion in the ferromagnetic bilayers of La$_{2-2x}$Sr$_{1+2x}$Mn$_2$O$_7$, $x = 0.30$–$0.40$ can be understood entirely within the double exchange picture, and (iii) the anomalously large low-temperature spin wave damping, which rises linearly with spin wave energy throughout the Brillouin zone, and contains no special 'resonances' where the damping might be enhanced due to crossing of other excitation branches such as phonons or orbitons.

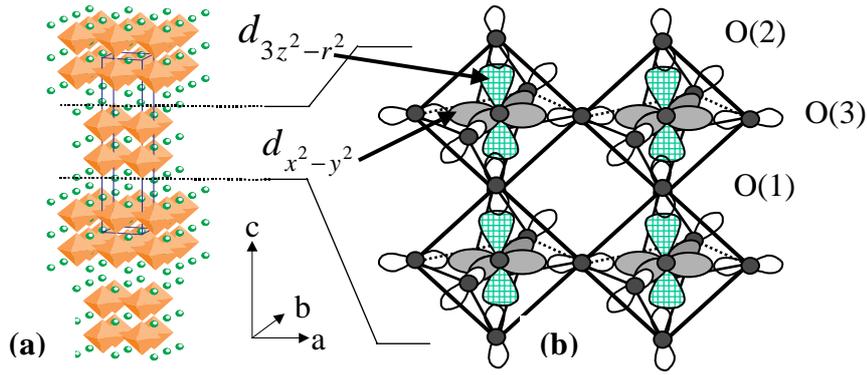

**Figure 1**
(a) Crystal structure of $La_{2-2x}Sr_{1+2x}Mn_2O_7$. The Mn ions are at the centre of the $MnO_6$ octahedra. Circles denote La and Sr. (b) Expanded view of $MnO_6$ octahedra in a single bilayer, showing the Mn $d_{3z^2-r^2}$ and $d_{x^2-y^2}$ orbitals.

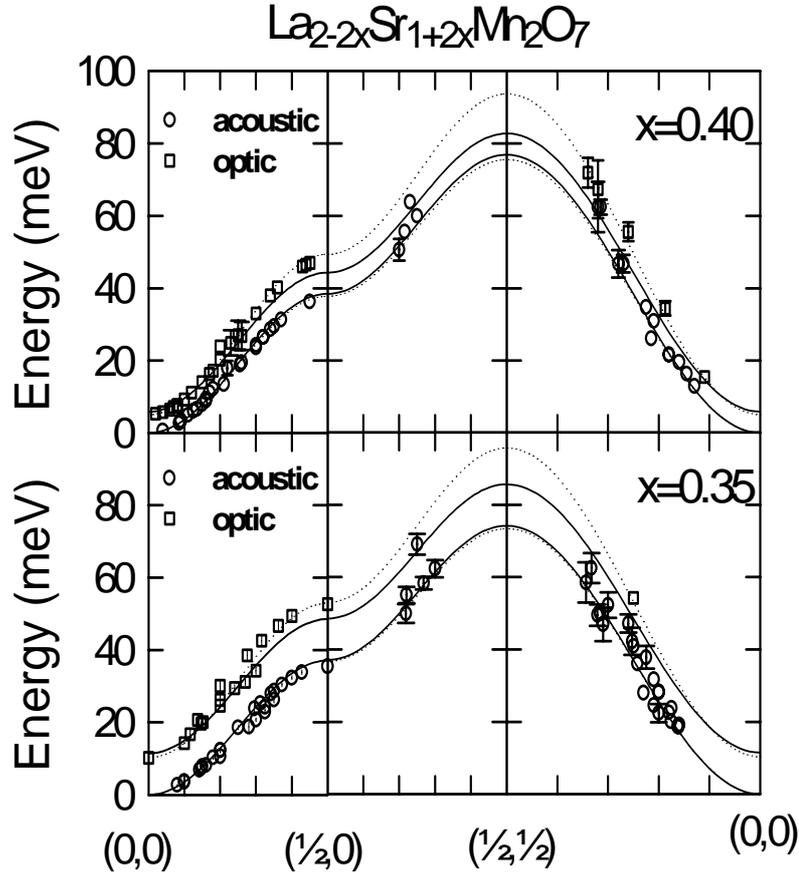

**Figure 2**
Measured spin wave dispersion relations for $x=0.40$, $x=0.35$. Solid lines show the best fits to the nnHFM model (parameters in text). Dashed lines show the best fit including diagonal exchange as described in the text. (For $x=0.40$, $SJ_\parallel = 10.3 \pm 0.1$, $SJ_\perp = 5.8 \pm 0.5$, $SJ_{a,0,a} = -0.8 \pm 0.1$ meV; for $x=0.35$, $SJ_\parallel = 9.9 \pm 0.2$, $SJ_\perp = 8.1 \pm 0.7$, $SJ_{a,0,a} = -0.8 \pm 0.2$ meV). Error bars are omitted if smaller than the markers.

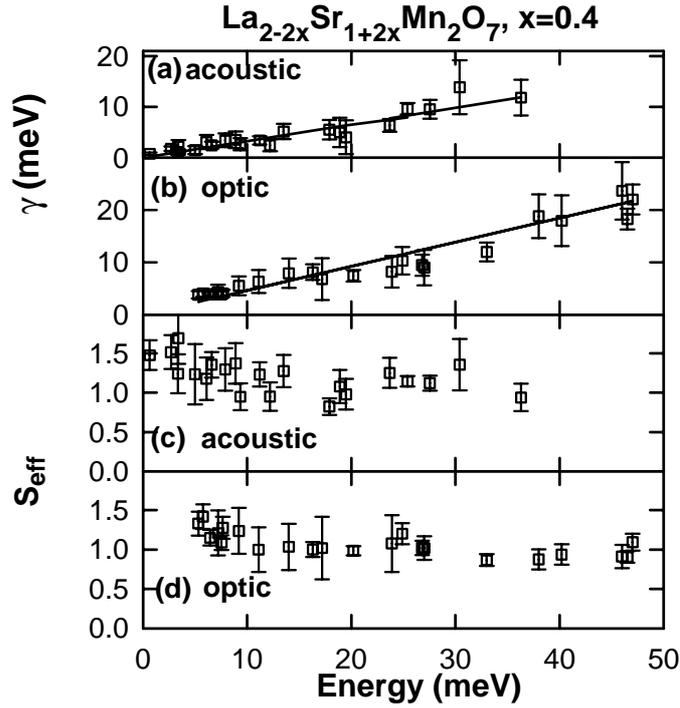

**Figure 3**

Energy dependence for $x=0.40$ of the inverse spin wave lifetime, $\gamma$, for (a) acoustic (b) optic spin waves, and the energy dependence of the spin wave intensity $S_{eff}$ in the scattering cross-section for the nnHFM model (see text) for (c) acoustic (d) optic spin waves. Error bars are omitted if smaller than the markers.

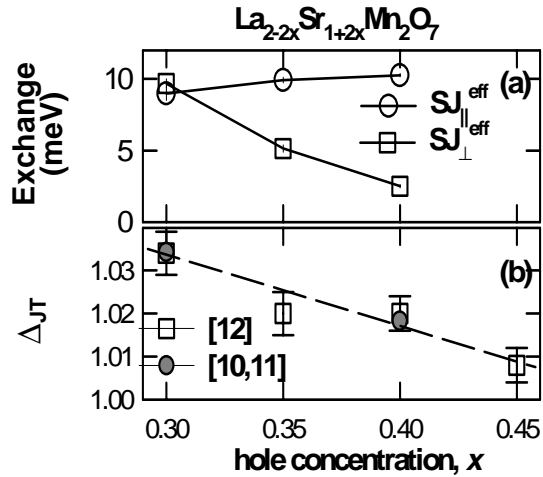

**Figure 4**

(a) $SJ_\parallel^{eff}$ and $SJ_\perp^{eff}$ for the nnHFM model derived from the spin wave stiffness and optic gap. For $x=0.35$ and 0.40, the data come from the fits to the dispersion relation. For $x=0.30$, only long wavelength data was taken, and $SJ_\parallel^{eff}$ and $SJ_\perp^{eff}$ are the result of fits directly to the counts. (b) Distortion of the $MnO_6$ octahedra, where $\Delta_{JT} = (1/2)(d_{Mn-O(1)} + d_{Mn-O(2)})/d_{Mn-O(3)}$. Error bars are omitted if smaller than the markers.